\documentclass[11pt,preprint,superscriptaddress,amsmath,amssymb,11pt,aps,showpacs,showkeys,nofootinbib,a4paper]{article}
\usepackage{amssymb,amsmath,amsfonts}
\usepackage{graphicx}
\usepackage{eepic,epsfig}
\usepackage{verbatim}
\usepackage{color}
\usepackage{hyperref}
1.60cm \makeatletter \@addtoreset{equation}{section}

\makeatother

\begin{document}
\title{Induced fermionic current in AdS spacetime in the presence of a cosmic string and a compactified dimension}
\author{S. Bellucci$^1$\thanks {E-mail: bellucci@lnf.infn.it}, \ W. Oliveira dos Santos$^{1,2}$\thanks {E-mail: wagner.physics@gmail.com} \ and E. R. Bezerra de Mello$^2$\thanks
{E-mail: emello@fisica.ufpb.br}\\
\\
\textit{$^1$INFN, Laboratori Nazionali di Frascati,}\\
\textit{Via Enrico Fermi 40, 00044 Frascati, Italy}\\
\textit{$^2$Departamento de F\'{\i}sica, Universidade Federal da Para\'{\i}ba,}\\
\textit{58059-900, Caixa Postal 5008, Jo\~{a}o Pessoa, PB, Brazil}\vspace{%
0.3cm}\\
}
\maketitle
%
\begin{abstract}
In this paper, we consider a massive charged fermionic quantum field and investigate the current densities induced by a magnetic flux running along the core of an idealized cosmic string in the background geometry of a 5-dimensional anti-de Sitter spacetime, assuming that an extra dimension is compactified. Along the compact dimension quasi-periodicity condition is imposed on the field with a general phase. Moreover, we admit the presence of a magnetic flux enclosed by the compactified axis. The latter gives rise to Ahanorov-Bohm-like effect on the vacuum expectation value of the currents. In this setup, only azimuthal and axial current densities take place. The former presents two contributions, with the first one due to the cosmic string in a 5-dimensional AdS spacetime without compact dimension, and the second one being induced by the compactification itself. The latter is an odd function of the magnetic flux along the cosmic string and an even function of the magnetic flux enclosed by the compactified axis with period equal to the quantum flux. As to the induced axial current, it is an even function of the magnetic flux along the string's core and an odd function of the magnetic flux enclosed by the compactification perimeter. For untwisted and twisted field along compact dimension, the axial current vanishes. The massless field case is presented as well as some asymptotic limits for the parameters of the model. 

\end{abstract}
\bigskip
%
\section{Introduction}
\label{Int}
In the context of grand unified theories, a consequence of the vacuum symmetry breaking phase transitions is the formation of different types of topological defects in the early Universe \cite{Kibble,V-S}. In particular, cosmic strings have been receiving considerable attention. Although the recent observational data on the cosmic microwave background (CMB) have ruled out cosmic strings as a primary source for primordial density perturbations, they are still possible candidates for the generation of a variety of interesting physical effects such as gamma ray bursts \cite{Berezinski}, high energy cosmic rays \cite{Bhattacharjee} and gravitational waves \cite{Damour}. In addition, cosmic strings have also been considered in scenarios beyond the standard model of particles, like in supersymmetry and in string theory approach \cite{Hindmarsh,Copeland}. More recently, the proposal of a formation mechanism for cosmic strings in the framework of brane-inflation has been attracting renewed interest on this type of topological defects \cite{Sarangi}.

The simplest theoretical model describes a cosmic string as an infinitely long and straight object which locally presents no curvature, except on its core, where it has a delta shaped curvature tensor \cite{Vile81}. Moreover, a cosmic string in this model is mainly characterized by a deficit angle on the two-dimensional sub-space perpendicular to its core. The corresponding nontrivial topology gives rise to a number of interesting physical phenomena. One of these concerns the effect of a cosmic string on the vacuum structure. 

The presence of a cosmic string in anti-de Sitter (AdS) spacetime provides an even more interesting framework, given that the combined geometry allows to distinguish the source of each part in the vacuum expectation values (VEV) of some observables. An interesting feature of the negative cosmological constant solution of the General Relativity field equations, is the fact that it makes it possible to solve several problems, exactly as a consequence of its high symmetry, allowing for easier quantization of the fields, besides providing deep insights into the quantization of the fields in other curved spacetimes. Furthermore, the AdS spacetime appears as a ground-state solution of string and supergravity theories and plays a fundamental role in the context of the AdS/CFT correspondence, a framework that realizes the holographic principle, relating the aforementioned fundamental theories with conformal field theories living in their boundaries \cite{aharony}. Moreover, the AdS space also appears in braneworlds scenarios with large extra dimensions, providing a way to overcome the hierarchy problem between the gravitational
and electroweak mass scales \cite{Brax}.

Another typical ingredient coming from the context of the above-mentioned fundamental theories, in which AdS spacetime plays a relevant role, is given by compact extra dimensions. The latter induce non-vanishing contributions to the expectation values of physical observables, such as the  energy-momentum tensor, which has not only the energy density component, but also the stress components (see \cite{deMello:2014hya} and references therein). In this case, for instance, the vacuum energy density induced by the extra compact dimensions offers an explanation for the observed and still unexplained accelerated expansion of the Universe at recent epoch. In Kaluza-Klein-type models and in braneworld scenarios, on the other hand, the dependence of the size of the compact extra dimension by the vacuum energy density serves as a mechanism to stabilize fields known as moduli fields \cite{P.Kanti_2002}.

All the above-mentioned ingredients can provide modifications in the vacuum quantum fluctuations of relativistic fields in curved spacetime. In particular, the induced VEV of the current density by curved backgrounds with a cosmic string has been investigated in Refs.\cite{deMello:2014hya,Bragana,Mello2015,Mello2013,Mello2010,Mello2010b,Souza}. In AdS spacetime with compact dimension, investigations of the vacuum polarizations as well as induced vacuum currents have been carried out in \cite{Santos:2018ttf,Mello2012,Bellucci2015jhep,Oliveira2019,Bellucci2016,Bellucci} providing strong motivations to study it.

In the present paper, we are mainly interested in investigating the effects on the vacuum current densities associated with a charged massive Dirac field arising from the geometry
and topology of a 5-dimensional AdS spacetime, in the presence of a cosmic string carrying a magnetic flux. Also we will admit the compactification of the extra dimension and the existence of a magnetic flux enclosed by the compactified axis. The non-trivial topology of this compactified extra dimension, as well as the magnetic fluxes, will give rise to additional contributions to the VEV of the current densities.

The paper is organized as follows. In the next section we obtain the mode-functions for a massive charged fermionic quantum field in the setup considered. In section \ref{sec3} these mode-functions are used for the evaluation of the induced azimuthal current density. By making use of the Abel-Plana summation formula, the 
latter is decomposed as the
sum of contributions induced by the cosmic string without compactification, and a second one induced by compactification itself. For these contributions we investigate some asymptotic limits of the model parameters. Towards the end of this section, we show that the charge density as well as the currents along the radial and Poincaré coordinates are zero. In section \ref{sec4} we calculate the axial current density and investigate its behavior for some specific asymptotic limits. Section \ref{sec5} is devoted to drawing the main conclusions about our results. Throughout the paper, we use natural units $G=\hbar=c=1$.
%
\section{Dirac Equation in (1+4)}
In this section we will obtain the fermionic wave-function associated to a massive charged quantum field, $\psi(x)$, in the background geometry of a $(1+4)$-dimensional AdS spacetime in the presence of a cosmic string and a compactified extra dimension. The mode-summation approach is used and the mode-functions will be needed in the calculation of the current densities.  

\subsection{Setup}
By using cylindrical coordinates system, the geometry associated with a cosmic string in a $(3+1)$-dimensional AdS spacetime is given by the line element below
\begin{equation}
ds^{2}=e^{-2y/a}[dt^{2}-dr^{2}-r^{2}d\phi ^{2}]-dy^{2}\ ,  \label{ds1}
\end{equation}
where $r\geqslant 0$ and $\phi \in \lbrack 0,\ 2\pi /q]$ define the
coordinates on the conical geometry, $(t, \ y)\in (-\infty ,\ \infty )$, and
the parameter $a$ determines the curvature scale of the background
spacetime. In the above coordinate system the string is along the $y-$axis.  The parameter $q\geq 1$ defines the planar angle deficit on the two-dimensional surface orthogonal to the string. Using the \textit{Poincar\'{e}} coordinate defined by $w=ae^{y/a}$, the line element above can be conformally related to the line element associated with a cosmic string in Minkowski spacetime
\begin{equation}
ds^2 = \left(\frac{a}{w}\right)^2[dt^2 - dr^2 - r^2d\phi^2 - dw^2 ] \ .
\label{ds2}
\end{equation}
For the new coordinate one has $w\in \lbrack 0,\ \infty )$. Specific values for this coordinates deserve to be mentioned: $w=0$ and $w=\infty $ correspond to the AdS boundary and horizon, respectively.

In order to write the line element (\ref{ds2}) in a five-dimensional AdS spacetime, we adopt the standard procedure, by adding an extra spatial coordinate \cite{deMello:2011ji}:
\begin{equation}
ds^2 = \left(\frac{a}{w}\right)^2\bigg[dt^2 - dr^2 - r^2d\phi^2 - dw^2 - dz^2\bigg]   \   .
\end{equation}
The cosmological constant, $\Lambda $, and the Ricci scalar, $R$, are related with the scale $a$ by the formulas
\begin{equation}
\Lambda =-\frac{6}{a^{2}} \ , \ \ R=-\frac{20}{a^{2}}\ .
\label{LamR}
\end{equation}

The dynamics of a spinorial quantum field in curved spacetime coupled with a gauge field, $A_{\mu}$, is given by the following equation:
\begin{equation}
	i\gamma^{\mu}(\nabla_{\mu}+ieA_{\mu})\psi-\tilde{s}m\psi=0, \quad \nabla_{\mu}=\partial_{\mu}+\Gamma_{\mu}, \quad \tilde{s}=\pm1 \ ,
	\label{Dirac-equation}
\end{equation}
where $\gamma^{\mu}$ are the Dirac matrices in curved spacetime and $\Gamma_{\mu}$ is the spin connection. Moreover, the two possible values of $\tilde{s}$ corresponds to the two irreducible representations of Dirac matrices in spacetimes with odd number of dimensions. Both sets of matrices in curved spacetime are related to the flat ones, $\gamma^{(a)}$, by the relations,
\begin{equation}
	\gamma_{\mu}=e^{\nu}_{(a)}\gamma^{(a)},\quad \Gamma_{\mu}=\frac{1}{4}\gamma^{(a)}\gamma^{(b)}e^{\nu}_{(a)}e_{(b)\nu,\mu}.
\end{equation}
The tetrad basis, $e^{\mu}_{(a)}$, satisfies the relation $e^{\mu}_{(a)}e^{\nu}_{(b)}\eta^{ab}=g^{\mu\nu}$, with $\eta^{ab}$ being the Minkowski spacetime metric tensor.

Note that we will also consider the presence of a constant vector potential along the extra compact dimension. This compactification is implemented by assuming that $z\in[0, \ L]$, and the matter field obeys the quasiperiodicity condition below,
\begin{equation}
\psi(t,r,\phi, w, z + L) = e^{2\pi i\beta}\psi(t,r,\phi, w, z),
\label{QPC}
\end{equation}
where $0\leq\beta\leq 1$. The special cases $\beta =0$ and $\beta =1/2$ correspond to the untwisted and twisted fields, respectively, along the $z$-direction. 

The set of Dirac matrices in flat spacetime assumed in this paper is the following:
\begin{equation}
	\gamma^{(0)}=-i
	\left( {\begin{array}{cc}
		0 & 1 \\
	    -1 & 0 \\
		\end{array} } \right), \quad \gamma^{(a)}=-i
	\left( {\begin{array}{cc}
	\sigma_{a} & 0 \\
	0 & -\sigma_{a} \\
	\end{array} } \right), \ \text{with} \  (a)=1,2,3, \quad \gamma^{(4)}=i\left( {\begin{array}{cc}
	0 & 1 \\
	1 & 0 \\
	\end{array} } \right),
\end{equation}
where $\sigma_{1},\sigma_{2},\sigma_{3}$ are the Pauli matrices. It is easy to verify that the above matrices obey the Clifford algebra, $\{\gamma^{(a)},\gamma^{(b)}\}=2\eta^{ab}$. We take the tetrad basis as follows:
\begin{equation}
	e^{\mu}_{(a)}=\frac{w}{a}\left( {\begin{array}{ccccc}
		1 & 0 & 0 & 0 & 0\\
	    0 & \cos(q\phi) & -\sin(q\phi)/r & 0 & 0\\
	    0 & \sin(q\phi) & \cos(q\phi)/r & 0 & 0\\
	    0 & 0 & 0 & 1 & 0\\
	    0 & 0 & 0 & 0 & 1\\
		\end{array} } \right)
\end{equation}
where the index $(a)$ identifies the rows  of the matrix. With this choice, the gamma matrices take the form
\begin{equation}
\gamma^{0}=\frac{w}{a}\gamma^{(0)} \ , \quad \gamma^{l}=\frac{w}{a}\left( {\begin{array}{cc}
	\sigma^{l} & 0 \\
	0 & -\sigma^{l}\\
	\end{array} } \right) \ , \quad
\gamma^{4}=i\frac{w}{a}\left( {\begin{array}{cc}
	0 & 1 \\
	1 & 0\\
	\end{array} } \right)
	\label{Dirac-matrices-curved}
\end{equation}
where we have introduced the $2 \times 2$ matrices for $l=(r,\phi,w)$
\begin{equation}
	\sigma^{r}=\left( {\begin{array}{cc}
		0 & e^{-iq\phi} \\
		e^{iq\phi} & 0\\
		\end{array} } \right) \ , \ \sigma^{\phi}=-\frac{i}{r}\left( {\begin{array}{cc}
		0 & e^{-iq\phi} \\
		-e^{iq\phi} & 0\\
		\end{array} } \right) \ , \ \sigma^{w}=\left( {\begin{array}{cc}
		1 & 0 \\
		0 & -1\\
		\end{array} } \right) \ .
	\label{Pauli-matrices-curved}
\end{equation}
For the spin connection we have
\begin{equation}
	\Gamma_{\mu}=\frac{1}{2a}\gamma^{(3)}\gamma_{\mu}+\frac{(1-q)}{2}\gamma^{(1)}\gamma^{(2)}\delta^{\phi}_{\mu} \ , \quad \Gamma_{z}=0,
\end{equation}
and its contraction with the Dirac matrices in curved spacetime is given by,
\begin{equation}
	\gamma^{\mu}\Gamma_{\mu}=-\frac{2}{w}\gamma^{w}+\frac{1-q}{2r}\gamma^{r}.
\end{equation}
Thus the Dirac equation takes the following form:
\begin{equation}
	\bigg(\gamma^{\mu}(\partial_{\mu}+ieA_{\mu})	-\frac{2}{w}\gamma^{w}+\frac{1-q}{2r}\gamma^{r}+i\tilde{s}m\bigg)\psi=0.
\end{equation}

As we have already mentioned, in this paper we want to consider the presence of a vector potential along the string's core and enclosed by the compact dimension. In this case we assume
\begin{eqnarray}
A_\mu=(0, \ 0, \ A_\phi, \ 0, \ A_z) \ .
\end{eqnarray}

For positive energy solutions, assuming the time-dependence of the eigenfunction in the form $e^{-iEt}$ and the boost symmetry in the coordinate $z$ represented by eigenfunctions in the form $e^{ik_z z}$, we may decompose the spinor field, $\psi$, as
\begin{eqnarray}
	\bigg(\sigma^{l}\partial_{l}-\frac{2}{w}\sigma^{w}+\frac{1-q}{2r}\sigma^{r}+ieA_{\phi}\sigma^{\phi}-\frac{\tilde{s}ma}{w}\bigg)\varphi-i[E+(k_{z}+eA_{z})]\chi&=&0, \nonumber \\
	\bigg(\sigma^{l}\partial_{l}-\frac{2}{w}\sigma^{w}+\frac{1-q}{2r}\sigma^{r}+ieA_{\phi}\sigma^{\phi}+\frac{\tilde{s}ma}{w}\bigg)\chi+i[(k_{z}+eA_{z})-E]\varphi&=&0 \ ,
	\label{dub-diff-eq}
\end{eqnarray}
  where $\varphi$ and $\chi$ are the upper and lower components of the spinor field, respectively.
Now taking the function $\chi$ from the second equation into the first one, we get the following second order differential equation for the $\varphi$:
\begin{eqnarray}
	&&\bigg\{\partial_{r}^{2}+\frac{1}{r}\partial_{r}+\frac{1}{r^2}\bigg[\partial_{\phi}+ieA_\phi-\frac{i(1-q)}{2}\sigma_{w}\bigg]^2+\partial_{w}^{2}-\frac{4}{w}\partial_w+\frac{6-(\tilde{s}ma)^2-(-1)^l\tilde{s}ma}{w^2}\nonumber\\&&+\big[E^2-\tilde{k}_z^2\big]\bigg\}\varphi=0 \ ,
	\label{diff-eq}
\end{eqnarray}
where we have introduced the notation $\tilde{k}_z=k_z+eA_z$.
In order to find the solution for \eqref{diff-eq}, we may use the following Ansatz, compatible with the cylindrical symmetry of the problem:
\begin{equation}
	\varphi=\left( {\begin{array}{c}
		C_{1}R_{1}(r)W_{1}(w)e^{iqn_{1}\phi} \\
		C_{2}R_{2}(r)W_{2}(w)e^{iqn_{2}\phi} \\
		\end{array} } \right) \ ,
	\label{upper-comp}
\end{equation}
with $C_{1}$ and $C_{2}$ being two arbitrary constants.
Plugging this Ansatz into \eqref{diff-eq}, we can see that the solution of the radial equation which is regular on the string is expressed in terms of the Bessel function of the first kind, $R_{l}=J_{\beta_{l}}(\lambda r)$, where its order is given by
\begin{equation}
	\beta_{1}=|q(n_{1}+\alpha)-(1-q)/2| \ \ , \ \ \beta_{2}=|q(n_{2}+\alpha)+(1-q)/2| \ ,
\end{equation}
with $n_{l}=0,\pm1, \pm2,....$ Note that we have introduced the notation $\alpha=eA_{\phi}/q=-\Phi_{\phi}/\Phi_{0}$, where $\Phi_{0}=2\pi/e$ is the quantum flux. As to the differential equation associated with the Poincaré coordinate, the general solution is given in terms of a linear combination of the functions $w^{5/2}J_{\nu_l}(pw)$ and $w^{5/2}Y_{\nu_l}(pw)$, where $Y_{\nu}(x)$ is the Neumann function and
\begin{equation}\nu_{l}=|\tilde{s}ma+(-1)^l/2|.
\label{rel-nu}
\end{equation}
Considering $ma\ge1/2$, the Neumann function must be excluded according to the normalizability conditions of the mode functions, and therefore, we shall adopt the solution
\begin{equation}
	W_{l}(w)=w^{5/2}J_{\nu_{l}}(pw) \ .
\end{equation}

The energy associated to the modes is given by
\begin{equation}
E=\sqrt{\lambda^2+p^2+\tilde{k}_{z}^2} \ .
\end{equation}
For sake of simplicity, from now on we shall assume the representation of the Clifford algebra corresponding to $\tilde{s}=1$.

In short notation, the solution for the upper component is written as
\begin{equation}
	\varphi_{l}=C_{l}w^{5/2}J_{\beta_{l}}(\lambda r)J_{\nu_{l}}(pw)e^{iqn_{l}\phi}.
	\label{upper-spinor}
\end{equation}
Taking \eqref{upper-spinor} into the first equation in \eqref{dub-diff-eq}, after some intermediate steps, we obtain
\begin{eqnarray}
	\chi=w^{5/2}\left( {\begin{array}{c}
		B_{1}J_{\beta_{1}}(\lambda r)J_{\nu_{1}}(pw)e^{iqn_{1}\phi} \\
		B_{2}J_{\beta_{2}}(\lambda r)J_{\nu_{2}}(pw)e^{iqn_{2}\phi} \\
		\end{array} } \right)
		\label{lower-spinor}
\end{eqnarray}	
with the relations
\begin{equation}
	n_{2}=n_{1}+1 \ , \ \beta_{2}=\beta_{1}+\epsilon_{n_{1}},
\end{equation}
where $\epsilon_{n}=1$ for $n>-\alpha$ and $\epsilon_{n}=-1$ for $n<-\alpha$. The coefficients $B_{1,2}$ in \eqref{lower-spinor} are given by
\begin{equation}
	B_{1}=\frac{i}{E+\tilde{k}_{z}}(pC_{1}-\epsilon_{n}\lambda C_{2}) \ , \ B_{2}=\frac{i}{E+\tilde{k}_{z}}(\epsilon_{n}\lambda C_{1}+pC_{2}) \ .
	\label{coeff-relations}
\end{equation}

We can see from the upper and lower spinor components given by \eqref{upper-spinor} and \eqref{lower-spinor}, respectively, that the wave function, $\psi$, is an eigenfunction of the total angular momentum projected along the direction of the string
\begin{equation}
	\hat{J}_{w}\psi=\bigg(-i\partial_{\phi}+\frac{q}{2}\Sigma^w\bigg)\psi \ , \ \Sigma^w=\left( {\begin{array}{cc}
		\sigma^{w} & 0 \\
		0 & \sigma^{w}\\
		\end{array} } \right) \ ,
\end{equation}
where
\begin{equation}
	j=n_{1}+1/2 \ , \ j=\pm 1/2, \pm 3/2,... \ .
\end{equation}

Note that the components of the fermionic wave function obtained have four coefficients and two equations relating them, given in \eqref{coeff-relations}. The normalization condition on the wave function yields an additional relation. Thus, one of the coefficients remains arbitrary and in order to determine this coefficient some additional condition should be imposed on the coefficients. The imposition of this condition comes from the fact that the quantum numbers $\sigma=(\lambda,p,j,k_{z})$ do not specify the fermionic wave function uniquely and some additional quantum number is required.

In order to specify the second constant we will require the following relation between the upper and lower components \cite{Bordag}:
\begin{equation}
	\chi_{1}=\kappa\varphi_{1}, \ \chi_{2}=-\varphi_{2}/\kappa.
	\label{c-coeff}
\end{equation}
From the expressions for the spinor components we find the eigenvalues of the parameter $\kappa$,
\begin{equation}
	\kappa=\kappa_{s}=\frac{-\tilde{k}_{z}+s\sqrt{\tilde{k}_{z}^{2}+p^2}}{p}, \ s=\pm1,
\end{equation}
and the relation
\begin{equation}
	C_{2}=-\frac{\epsilon_{n}\kappa_{s}}{\lambda}\big(E-s\sqrt{\tilde{k}_{z}^2+p^2}\big)C_1,
\end{equation}
for the coefficients in \eqref{upper-spinor}. Note that now the fermionic wave function is uniquely specified by the set of quantum numbers $\sigma=(\lambda,p,j,k_{z},s)$. The eigenvalues of the quantum number $k_z$ are determined from the quasi-periodicity condition \eqref{QPC}
\begin{equation}
	k_z=k_l=2\pi(l+\beta)/L \ , \ \text{with} \ \ l=0,\pm1, \pm2,... \ .
\end{equation}

On the basis of all considerations above, the positive-energy fermionic wave function can be expressed as
\begin{equation}
	\psi^{(+)}_{\sigma}(x)=C^{(+)}_{\sigma}e^{-iEt+ik_{l}z}w^{5/2}\left( {\begin{array}{c}
		J_{\beta_{j}}(\lambda r)J_{\nu_{1}}(pw) \\
		-\epsilon_{j}\kappa_{s}b^{(+)}_{s}J_{\beta_{j}+\epsilon_{j}}(\lambda r)J_{\nu_{2}}(pw)e^{iq\phi} \\
		i\kappa_{s}J_{\beta_{j}}(\lambda r)J_{\nu_{2}}(pw)\\
		i\epsilon_{j}b^{(+)}_{s}J_{\beta_{j}+\epsilon_{j}}(\lambda r)J_{\nu_{1}}(pw)e^{iq\phi}
		\end{array} } \right)e^{iq(j-1/2)\phi} \ ,
	\label{positive-energy-wfunc}
\end{equation}
where $\epsilon_{j}=1$ for $j>-\alpha$ and $\epsilon_{j}=-1$ for $j<-\alpha$, and the order of the Bessel function is defined as
\begin{equation}
	\beta_{j}=q|j+\alpha|-\epsilon_{j}/2.
\end{equation}
The energy is expressed in terms of $\lambda$, $p$ and $\tilde{k}_{l}$ by the relation
\begin{equation}
	E=\sqrt{\lambda^2+p^2+\tilde{k}_{l}^2} \ ,
\end{equation}
where $\tilde{k}_{l}=2\pi (l+\tilde{\beta})/L$, with
\begin{equation}
	\tilde{\beta}=\beta+eA_{z}L/(2\pi)=\beta-\Phi_{z}/\Phi_{0}.
\end{equation}
In \eqref{positive-energy-wfunc} and in what follows below, we use the notation
\begin{equation}
	b^{(\pm)}_{s}=\frac{E\mp s\sqrt{\tilde{k}_{l}^2+p^2}}{\lambda}.
\end{equation}
Note that one has the relation $b^{(+)}_{s}b^{(-)}_{s}=1$.

The coefficient $C^{(+)}_{\sigma}$ in \eqref{positive-energy-wfunc} is determined from the normalization condition
\begin{equation}
	\int d^3x\sqrt{\gamma}(\psi^{(+)}_{\sigma})^\dagger\psi^{(+)}_{\sigma^\prime}=\delta_{\sigma,\sigma^\prime} \ , \
	\label{norm-cond}
\end{equation}
where $\gamma$ is the determinant of the spatial metric. The delta symbol on the right-hand side is understood
as the Dirac delta function for continuous quantum numbers $(\lambda, p)$ and the Kronecker delta for discrete
ones $(j,s,l)$. Taking the eigenspinor in \eqref{positive-energy-wfunc} into \eqref{norm-cond} and using the value of the standard integral involving the products of the Bessel functions \cite{Grad}, we find
\begin{equation}
	|C^{(+)}_{\sigma}|^{2}=\frac{sqp^2\lambda^2}{8\pi a^4LE\kappa_{s}b^{(+)}_{s}\sqrt{\tilde{k}_{l}^2+p^2}} \ .
\end{equation}

The negative-energy fermionic mode function can be obtained in a similar way. The corresponding result is given by the expression
\begin{equation}
\psi^{(-)}_{\sigma}(x)=C^{(-)}_{\sigma}e^{iEt+ik_{l}z}w^{5/2}\left( {\begin{array}{c}
	J_{\beta_{j}}(\lambda r)J_{\nu_{1}}(pw) \\
	\epsilon_{j}\kappa_sb^{(-)}_{s}J_{\beta_{j}+\epsilon_{j}}(\lambda r)J_{\nu_{2}}(pw)e^{iq\phi} \\
	i\kappa_{s}J_{\beta_{j}}(\lambda r)J_{\nu_{2}}(pw)\\
	-i\epsilon_{j}b^{(-)}_{s}J_{\beta_{j}+\epsilon_{j}}(\lambda r)J_{\nu_{1}}(pw)e^{iq\phi}
	\end{array} } \right)e^{iq(j-1/2)\phi} \ ,
\label{negative-energy-wfunc}
\end{equation}
and the normalization constant is given by the relation
\begin{equation}
	|C^{(-)}_{\sigma}|^{2}=\frac{sqp^2\lambda^2}{8\pi a^4LE\kappa_{s}b^{(-)}_{s}\sqrt{\tilde{k}_{l}^2+p^2}} \ .
\end{equation}
\section{Fermionic Current Density}
\label{sec3}
The VEV of the fermionic current density, $\langle j^\mu\rangle=e\bar{\psi}\gamma^\mu\psi$, can be evaluated by using the mode sum formula,
\begin{equation}
	\langle j^\mu\rangle=\frac{e}{2}\sum_{\sigma}[\bar{\psi}^{(-)}_\sigma\gamma^\mu\psi^{(-)}_\sigma-\bar{\psi}^{(+)}_\sigma\gamma^\mu\psi^{(+)}_\sigma] \ .
	\label{curr-dens-formula}
\end{equation}
where $\bar{\psi}_{\sigma}^{(\pm)}=\psi_{\sigma}^{(\pm)\dagger}\gamma^{(0)}$ is the Dirac adjoint, and the summation over $\sigma$ is a compact notation defined below,
\begin{equation}
	\sum_{\sigma}=\int_{0}^{\infty}d\lambda\int_{0}^{\infty}dp\sum_{s=\pm1}\sum_{j=\pm1/2,...}\sum_{l=-\infty}^{\infty} \ .
\end{equation}
As we will see, this VEV is a periodic function of the magnetic fluxes $\Phi_{\phi}$ and $\Phi_{z}$ with the period equal to the quantum flux. Thus, it is convenient to express the parameter $\alpha$ as
\begin{eqnarray}
	\alpha=n_0+\alpha_0,
\end{eqnarray}
where $n_0$ is an integer number and $|\alpha_0|<1/2$.
\subsection{Charge density, current densities in the radial and w-directions}
\label{subsec3.1}

Let us compute the VEV of the charge density
\begin{equation}
	\rho(x)=\langle j^0(x)\rangle=\frac{ew}{2a}\sum_{\sigma}[\psi^{(-)\dagger}_\sigma\psi^{(-)}_\sigma -\psi^{(+)\dagger}_\sigma\psi^{(+)}_\sigma]\ .
	\label{charge-density}
\end{equation}
Taking \eqref{positive-energy-wfunc} and \eqref{negative-energy-wfunc} into \eqref{charge-density}, we arrive directly to a vanishing charge density. Thus, there is no charge density induced.

As to the VEV of the radial current, substituting the mode functions for positive and negative-energy as well the Dirac matrices given in \eqref{Dirac-matrices-curved} and \eqref{Pauli-matrices-curved} into \eqref{curr-dens-formula}, we can see that there appears a cancellation between the terms. Thus, there is no induced radial current. A similar analysis also leads to a vanishing induced current density along the coordinate $w$.
\subsection{Azimuthal Current}
\label{subsec3.2}
The VEV of the azimuthal current is given by
\begin{equation}
\langle j^\phi\rangle=\frac{e}{2}\sum_{\sigma}[\bar{\psi}^{(-)}_\sigma\gamma^\phi\psi^{(-)}_\sigma-\bar{\psi}^{(+)}_\sigma\gamma^\phi\psi^{(+)}_\sigma] \ .
\label{sum-mode-formula}
\end{equation}
Substituting the positive and negative wave functions into the above expression, we get
\begin{eqnarray}
	&&\langle j^\phi\rangle=-\frac{qew^6}{2\pi La^5r}\int_{0}^{\infty}d\lambda\int_{0}^{\infty}dp\sum_{j
	}\epsilon_{j}p\lambda^2 J_{\beta_j}(\lambda r)J_{\beta_j+\epsilon_{j}}(\lambda r)[J^2_{\nu_2}(pw)+J^2_{\nu_1}(pw)]\nonumber \\&&\times\sum_{l=-\infty}^{\infty}\frac{1}{\sqrt{\lambda^2+p^2+[2\pi (l+\tilde{\beta})/L]^2}} \ ,
	\label{curr-phi-eq3}
\end{eqnarray}
where the sum over $s$ has been already performed and it has yielded an overall factor of 2.

In order to sum over the quantum number $l$, we shall apply the Abel-Plana formula in the form \cite{Saharian2010}
\begin{eqnarray}
	&&\sum_{l=-\infty}^{\infty}g(l+\tilde{\beta})f(|l+\tilde{\beta}|)=\int_{0}^{\infty}du[g(u)-g(-u)]f(u)\nonumber \\&&+i\int_{0}^{\infty}du[f(iu)-f(-iu)]\sum_{n=\pm1}\frac{g(i\lambda u)}{e^{2\pi(u+in\tilde{\beta})}-1} \ ,
	\label{abel-plana}
\end{eqnarray}
taking $g(u)=1$ and
\begin{equation}
	f(u)=\frac{1}{\sqrt{(2\pi u/L)^2+\lambda^2+p^2}} \ .
	\label{fu-function}
\end{equation}
As a result we may decompose the induced azimuthal current as
\begin{equation}
	\langle j^\phi\rangle=\langle j^\phi\rangle_{\rm{cs}}+\langle j^\phi\rangle_{\rm{c}}
	\label{VEV-phi-comp}
\end{equation}
where the first term, $\langle j^\phi\rangle_{\rm{cs}}$, comes from the first integral on the right-hand side of \eqref{VEV-phi-comp} and corresponds to the induced azimuthal current density in the geometry of a cosmic string without compactification. The second contribution, $\langle j^\phi\rangle_{\rm{c}}$, is induced by the compactification. As we shall see the latter vanishes in the limit $L\rightarrow\infty$.

Combining \eqref{curr-phi-eq3} and \eqref{abel-plana}, we get
\begin{eqnarray}
	\langle j^\phi\rangle_{\rm{cs}}&=&-\frac{qew^6}{2\pi^2 a^5r}\int_{0}^{\infty}d\lambda\int_{0}^{\infty}dp\sum_{j
	}\epsilon_{j}p\lambda^2 J_{\beta_j}(\lambda r)J_{\beta_j+\epsilon_{j}}(\lambda r)[J^2_{\nu_2}(pw)+J^2_{\nu_1}(pw)] \nonumber \\
	&\times&\int_{0}^{\infty}\frac{dx}{\sqrt{\lambda^2+p^2+x^2}} \ ,
\end{eqnarray}
where we have introduced the variable $x=2\pi u/L$.
Using the following integral representation:
\begin{equation}
	\frac{1}{\sqrt{\lambda^2+p^2+x^2}}=\frac{2}{\sqrt{\pi}}\int_{0}^{\infty}dse^{-s^2(\lambda^2+p^2+x^2)} \ ,
\end{equation}
we can carry out the integrations over all variables except over $s$, obtaining
	\begin{eqnarray}
	\langle j^\phi\rangle_{\rm{cs}}&=&-\frac{qew^6}{4\pi^2 a^5r^6}\int_{0}^{\infty}dyy^2e^{-(1+w^2/r^2)y}\bigg[J^2_{\nu_2}\bigg(\frac{w^2}{r^2}y\bigg)+J^2_{\nu_1}\bigg(\frac{w^2}{r^2}y\bigg)\bigg]\nonumber \\
	&\times&\big[\mathcal{I}(q,\alpha_0,y)-\mathcal{I}(q,-\alpha_0,y)\big] \ ,
	\label{VEV-n}
	\end{eqnarray}
	where we have introduced a new variable $y=r^2/2s^2$ and the function $\mathcal{I}(q,\alpha_0,y)$ is given by \cite{Mello}
	\begin{eqnarray}
	\mathcal{I}(q,\alpha_0,y)&=&\frac{e^y}{q}-\frac{1}{\pi}\int_{0}^{\infty}dz\frac{e^{-y\cosh y}f(q,\alpha_0,z)}{\cosh(qz)-\cos(q\pi)}\nonumber\\&+&\frac{2}{q}\sum_{k=1}^{p}(-1)^k\cos[2\pi k(\alpha_0-1/2q)]e^{y\cos(2\pi k/q)} \ ,
	\label{j-sum}
	\end{eqnarray}
	with $2p<q<2p+2$ and with the notation
	\begin{eqnarray}
		f(q,\alpha_0,z)&=&\cos[q\pi(1/2-\alpha_0)]\cosh[(q\alpha_0+q/2-1/2)z]\nonumber\\&-&\cos[q\pi(1/2+\alpha_0)]\cosh[(q\alpha_0-q/2-1/2)z] \ .
		\label{f-function}
	\end{eqnarray}
	Using the formula \eqref{j-sum}, after the integration over $y$ with the help of \cite{Grad}, \eqref{VEV-n} can be presented in the form
	\begin{eqnarray}
		\langle j^\phi\rangle_{\rm{cs}}&=&-\frac{8e}{(2\pi)^{5/2} a^5}\bigg[\sideset{}{'}{\sum}_{k=1}^{[q/2]}(-1)^k\sin(\pi k/q)\sin(2\pi\alpha_0 k)\mathcal{Z}(u_k)\nonumber \\
		&+&\frac{q}{\pi}\int_{0}^{\infty}dz\frac{\cosh(z)g(q,\alpha_0,2z)}{\cosh(2qz)-\cos(q\pi)}\mathcal{Z}(u_z)\bigg] \ ,
		\label{j-phi-cs}
	\end{eqnarray}
	where $[q/2]$ represents the integer part of $q/2$ and the prime
	on the sign of the summation means that in the case $q=2p$
	the term $k=q/2$ should be taken with the coefficient $1/2$. Moreover, the function $g(q,\alpha_0,z)$ is defined as
	\begin{eqnarray}
	g(q,\alpha_0,z)&=&\cos[q\pi(1/2+\alpha_0)]\cosh[q(1/2-\alpha_0)z]\nonumber\\&-&\cos[q\pi(1/2-\alpha_0)]\cosh[q(1/2+\alpha_0)z] \ .
	\label{g-function}
	\end{eqnarray}
	In \eqref{j-phi-cs} we have introduced the function
	\begin{equation}
		\mathcal{Z}(u)=\sum_{i=1,2}F_{\nu_i}(u) \ , \ \text{with}  \ F_{\nu_i}=\frac{e^{-i5\pi/2}Q^{5/2}_{\nu_i-1/2}(u)}{(u^2-1)^{5/4}} \ ,
		\label{function-z}
	\end{equation}
	where $Q_\mu^\nu(z)$ represents the Legendre Associated Functions of second kind \cite{Grad}; moreover, we have defined the new variables
\begin{eqnarray}
u_{k}&=&1+2(r/w)^2\sin^2(\pi k/q) \ , \nonumber \\ u_{z}&=&1+2(r/w)^2\cosh^2(z) \ .
\label{var}
\end{eqnarray}
Let us now consider some special cases of the function $\mathcal{Z}(u)$ defined in \eqref{function-z}. For values of  $ma\ge1/2$, according to \eqref{rel-nu}, we have the relation $\nu_2=\nu_1+1$. In this case, for large values of the argument, $u\gg1$, the function $\mathcal{Z}(u)$ is simplified and to the leading term takes the form
\begin{eqnarray}
	\mathcal{Z}(u)\approx\frac{\sqrt{2\pi}(1+\nu_2)\nu_2}{2^{\nu_2}u^{\nu_2+2}} \ .
	\label{function-z-gg-1}
\end{eqnarray}
Now considering $m=0$, it implies from \eqref{rel-nu} that $\nu_1=\nu_2=1/2$. In this case, simplifications can be carried out and equation \eqref{function-z} may be presented as
\begin{eqnarray}
	\mathcal{Z}(u)=\frac{3\sqrt{\pi}}{4}\frac{1}{(u-1)^{5/2}} \ .
	\label{function-z-m0}
\end{eqnarray}

\begin{figure}[!htb]
	\begin{center}
		\centering
		\includegraphics[scale=0.45]{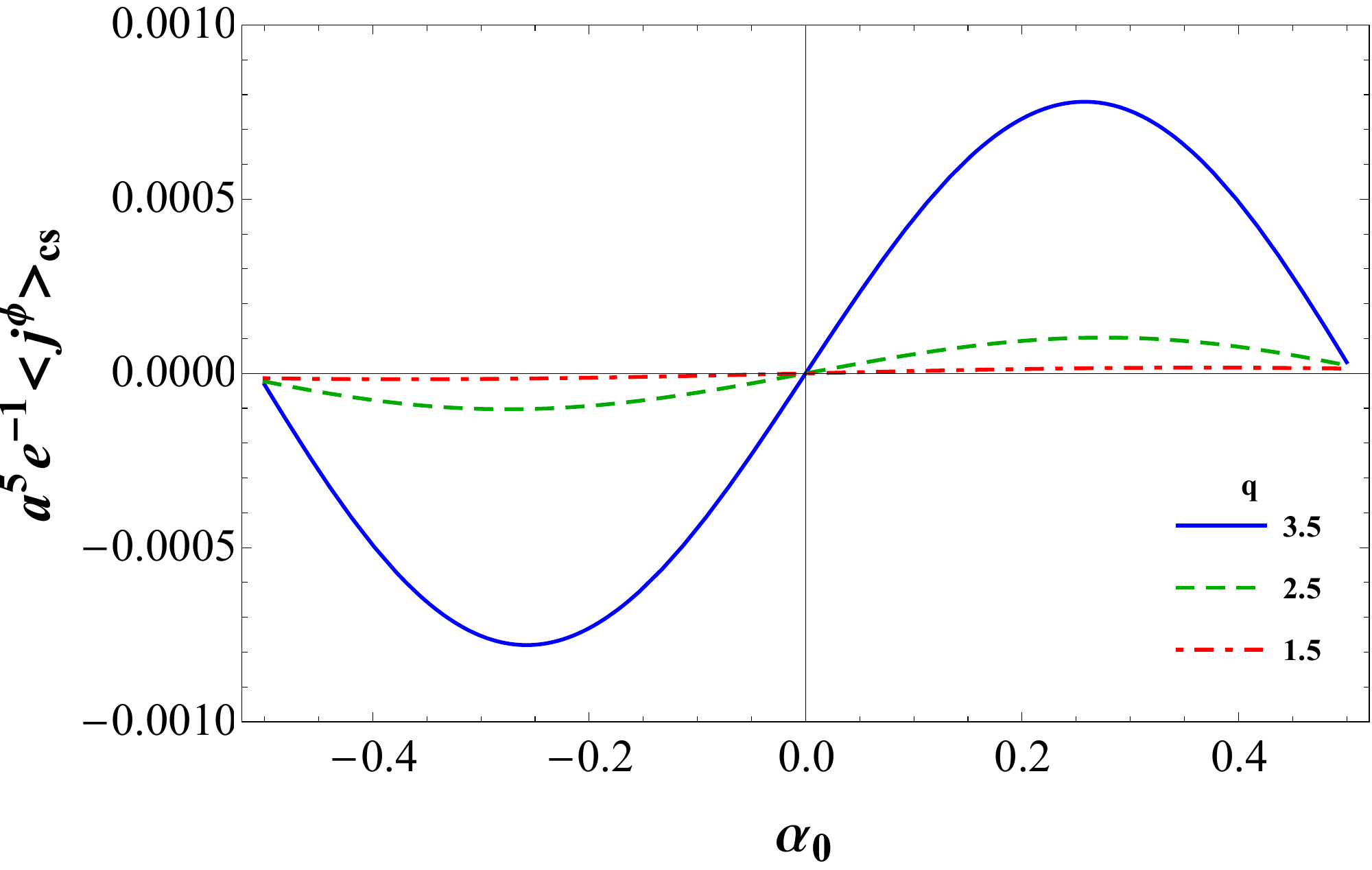}
		\caption{The VEV of the induced azimuthal current without compactification in \eqref{j-phi-cs} is plotted, in units of $a^{5}e^{-1}$, as a function of $\alpha_0$ for $r/w=1$, $ma=5$, $q=1.5,2.5$ and $3.5$.}
		\label{fig1}
	\end{center}
\end{figure}

In Fig.\ref{fig1} we plot the string contribution to the azimuthal current as a function of the magnetic flux along the string for $q=1.5,2.5$ and $3.5$. As we can see, $\langle j^\phi\rangle_{\rm{cs}}$ is an odd function of $\alpha_0$ and has its amplitude depending on the deficit angle caused by the presence of the string.
\\

\begin{figure}[!htb]
	\begin{center}
		\centering
		\includegraphics[scale=0.45]{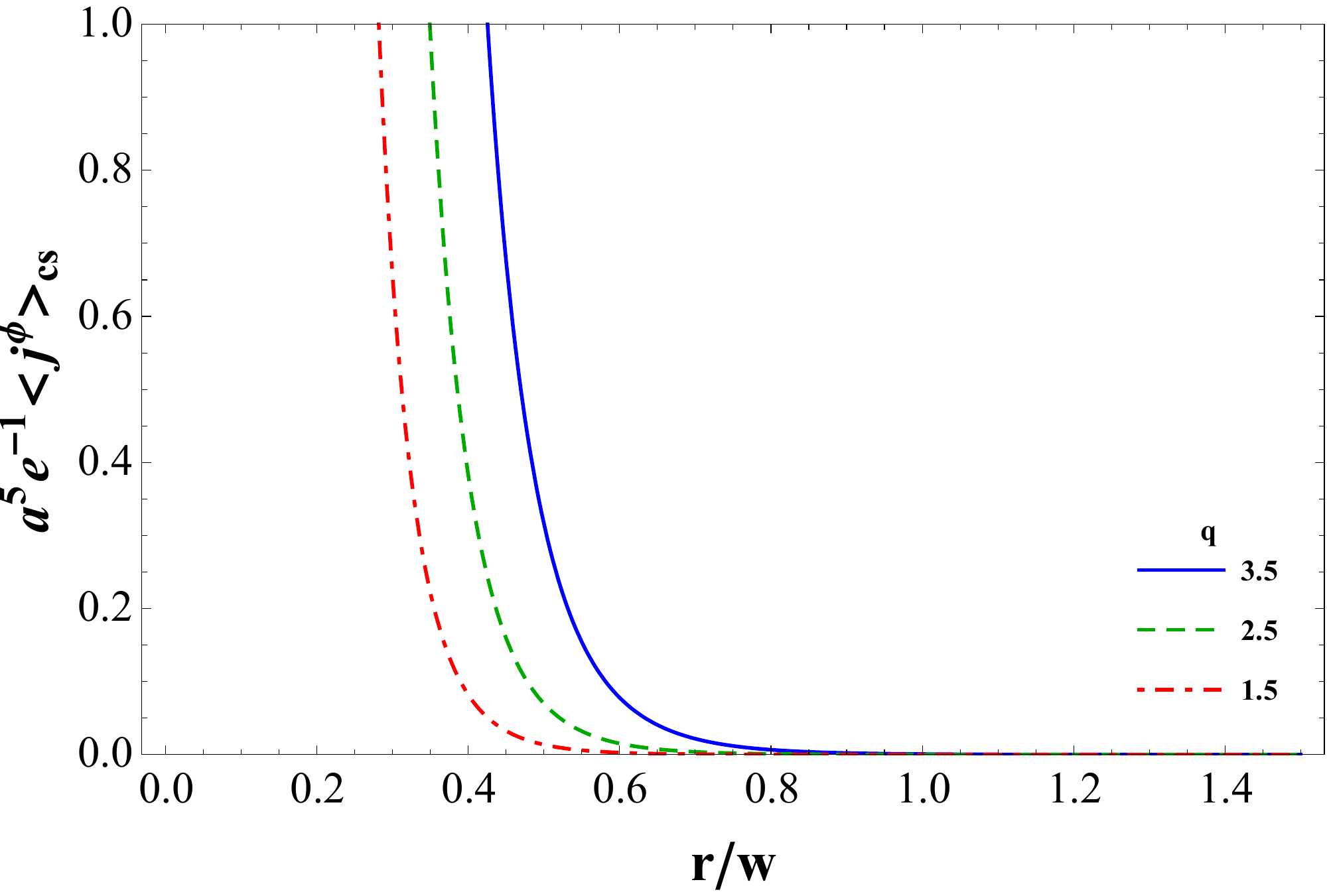}
		\caption{The VEV of the string contribution to the induced azimuthal current in Eq.\eqref{j-phi-cs} is plotted, in units of $a^{5}e^{-1}$, as a function of $r/w$ for $\alpha_0=0.25$, $ma=5$, $q=1.5,2.5$ and $3.5$.}
		\label{fig2}
	\end{center}
\end{figure}

Now we turn our attention to the investigation of \eqref{j-phi-cs} for small and large values of the ratio $r/w$ by using the asymptotic expressions from Ref.\cite{deMello:2014hya}. For points close to the string, $r/w\ll1$, we have the following expression:
\begin{eqnarray}
	\langle j^\phi\rangle_{\rm{cs}}\approx-\frac{3e}{(4\pi)^2 }\bigg(\frac{w}{ar}\bigg)^5\Bigg[\sideset{}{'}{\sum}_{k=1}^{[q/2]}(-1)^k\frac{\sin(2\pi\alpha_0 k)}{\sin^4(\pi k/q)}
	+\frac{q}{\pi}\int_{0}^{\infty}dz\frac{\cosh^{-4}(z)g(q,\alpha_0,2z)}{\cosh(2qz)-\cos(q\pi)}\Bigg] \ .
	\label{asymp-close-cs}
\end{eqnarray}
From the above expression we can see that this VEV diverges as $(w/r)^5$.
On the other hand, for points far away from the string, $r/w\gg1$, and values of $ma\ge1/2$, we find by plugging \eqref{function-z-gg-1} into \eqref{j-phi-cs}, the following asymptotic expression:
\begin{eqnarray}
	\langle j^\phi\rangle_{\rm{cs}}&\approx&-\frac{\nu_2(1+\nu_2) e}{2^{2\nu_2+1}\pi^2 a^5}\bigg(\frac{w}{r}\bigg)^{2\nu_2+4}\Bigg[\sideset{}{'}{\sum}_{k=1}^{[q/2]}(-1)^k\frac{\sin(2\pi\alpha_0 k)}{\sin^{2\nu_2+3}(\pi k/q)}
	\nonumber\\&+&\frac{q}{\pi}\int_{0}^{\infty}dz\frac{\cosh^{-(2\nu_2+3)}(z)g(q,\alpha_0,2z)}{\cosh(2qz)-\cos(q\pi)}\Bigg] \ .
	\label{asymp-far-away-cs}
\end{eqnarray}
We can see from this expression that in this regime the cosmic string influence on the intensity of the azimuthal current starts to fade and the decay behavior depends on the mass of the particle. In Fig.\ref{fig2} is displayed the string contribution without compactification as a function of the proper distance in units of $a$, $r/w$, for different values of $q$. This plot shows that this contribution diverges on the string, $r=0$, as well as it vanishes for distant points with its decay behavior depending on $q$. 
\\

\begin{figure}[!htb]
	\begin{center}
		\centering
		\includegraphics[scale=0.5]{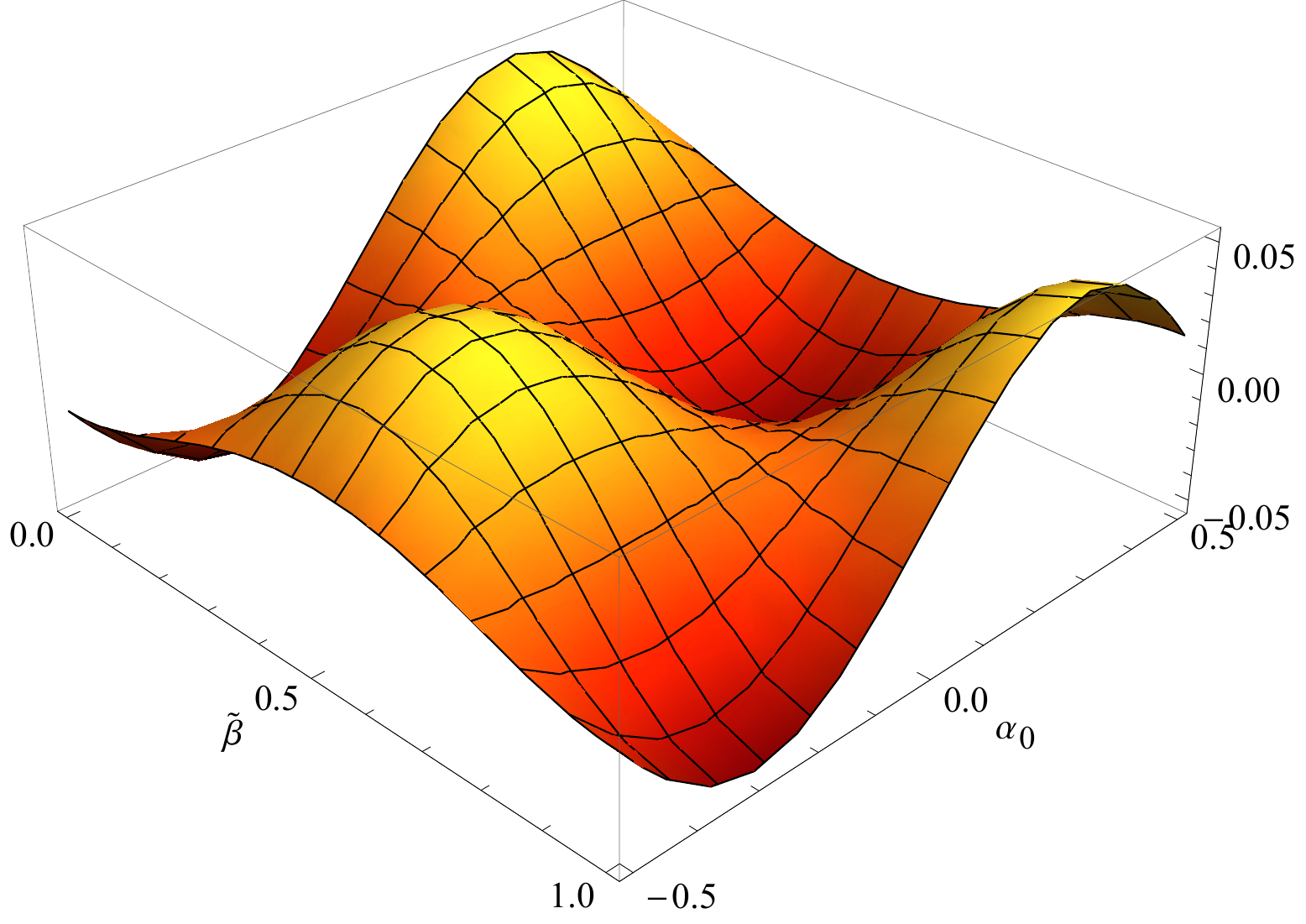}
		\caption{The azimuthal current induced by compactification given in \eqref{j-phy-c} is plotted in units of $a^5e^{-1}$ as a function of $\tilde{\beta}$ and $\alpha_0$ for fixed $r/w=1$, $ma=5$ and $q=2.5$.}
		\label{fig3}
	\end{center}
\end{figure}

Now let us develop the contribution induced by the compactification
\begin{eqnarray}
	\langle j^\phi\rangle_{\rm{c}}&=&-\frac{ieqw^6}{2\pi^2a^5r}\int_{0}^{\infty}d\lambda\int_{0}^{\infty} dp \sum_{j}\epsilon_j p\lambda^2J_{\beta_j}(\lambda r)J_{\beta_j+\epsilon_j}(\lambda r)[J^2_{\nu_{1}}(pw)+J^2_{\nu_{2}}(pw)] \nonumber \\
	&\times&\int_{\sqrt{\lambda^2+p^2}}^{\infty}\frac{dx}{\sqrt{x^2-\lambda^2-p^2}}\bigg(\frac{1}{e^{Lx+2\pi i\tilde{\beta}}-1}+\frac{1}{e^{Lx-2\pi i\tilde{\beta}}-1}\bigg) \ .
	\label{VEV-compact}
\end{eqnarray}
In order to proceed with the calculation, we shall use the series expansion, $(e^u-1)^{-1}=\sum_{l=1}^{\infty}e^{-lu}$ in the above expression. Taking this expansion into \eqref{VEV-compact}, we can perform the integral $x$ with the help of \cite{Grad}. After some few intermediate steps, we obtain
\begin{eqnarray}
	\langle j^\phi\rangle_{\rm{c}}&=&-\frac{eqw^6}{\pi^2a^5r}\sum_{l=1}^{\infty}\cos(2\pi l\tilde{\beta})\sum_{j}\epsilon_{j}\int_{0}^{\infty}d\lambda\lambda^2J_{\beta_j}(\lambda r)J_{\beta_j+\epsilon_j}(\lambda r)\nonumber \\
	&\times&\int_{0}^{\infty} dp p[J^2_{\nu_{1}}(pw)+J^2_{\nu_{2}}(pw)]K_0(lL\sqrt{\lambda^2+p^2}) \ .
	\label{j-compact-eq-inter-form}
\end{eqnarray} 
Using the following integral representation for the Macdonald function \cite{Abra}:
\begin{equation}
	K_{\nu}(x)=\frac{1}{2}\bigg(\frac{x}{2}\bigg)^\nu\int_{0}^{\infty}ds\frac{e^{-s-x^2/4s}}{s^{\nu+1}} \ ,
	\label{Macdonald-int-represent}
\end{equation}
we may write \eqref{j-compact-eq-inter-form} as
\begin{eqnarray}
	\langle j^\phi\rangle_{\rm{c}}&=&-\frac{eqw^6}{2\pi^2a^5r}\sum_{l=1}^{\infty}\cos(2\pi l\tilde{\beta})\int_{0}^{\infty}ds\frac{e^{-s}}{s}\sum_{j}\epsilon_{j}\int_{0}^{\infty}d\lambda\lambda^2J_{\beta_j}(\lambda r)J_{\beta_j+\epsilon_j}(\lambda r)e^{-l^2L^2\lambda^2/(4s)}\nonumber \\
	&\times&\int_{0}^{\infty} dp p[J^2_{\nu_{1}}(pw)+J^2_{\nu_{2}}(pw)]e^{-l^2L^2p^2/(4s)} \ \ .
\end{eqnarray}
Performing the integrations over $\lambda$ and $p$ with the help of \cite{Grad}, we get
\begin{eqnarray}
\langle j^\phi\rangle_{\rm{c}}&=&-\frac{eqw^6}{2\pi^2a^5r^6}\sum_{l=1}^{\infty}\cos(2\pi l\tilde{\beta})\int_{0}^{\infty}dyy^2e^{-[1+w^2/r^2+(lL)^2/2r^2]y}\nonumber\\&\times&\bigg[I_{\nu_1}\bigg(\frac{w^2}{r^2}y\bigg)+I_{\nu_2}\bigg(\frac{w^2}{r^2}y\bigg)\bigg][\mathcal{I}(q,\alpha_0,y)-\mathcal{I}(q,-\alpha_0,y)] \ ,
\end{eqnarray}
where we have introduced the variable $y=2sr^2/(lL)^2$.
Using the representation given in \eqref{j-sum} for the function $\mathcal{I}(q,\pm\alpha_0,y)$, we can perform the integral over $y$, arriving to the expression
\begin{eqnarray}
	\langle j^\phi\rangle_{\rm{c}}&=&-\frac{16e}{(2\pi)^{5/2}a^5}\sum_{l=1}^{\infty}\cos(2\pi l\tilde{\beta})\bigg[\sideset{}{'}{\sum}_{k=1}^{[q/2]}(-1)^k\sin(\pi k/q)\sin(2\pi\alpha_0 k)\mathcal{Z}(u_{lk})\nonumber \\
	&+&\frac{q}{\pi}\int_{0}^{\infty}dz\frac{\cosh(z)g(q,\alpha_0,2z)}{\cosh(2qz)-\cos(q\pi)}\mathcal{Z}(u_{lz})\bigg] \ ,
	\label{j-phy-c}
\end{eqnarray}
where we have introduced the variables
\begin{eqnarray}
	u_{lk}&=&1+\frac{(lL)^2+4r^2\sin^2(\pi k/q)}{2w^2} \ , \nonumber \\ u_{lz}&=&1+\frac{(lL)^2+4r^2\cosh^2(z)}{2w^2} \ .
	\label{var2}
\end{eqnarray}
In Fig.\ref{fig3}, $\langle j^\phi\rangle_{\rm{c}}$ is plotted in units of $a^5e^{-1}$ as a function of $\alpha_0$ and $\tilde{\beta}$, for $ma=5$ and $q=2.5$. From this plot we can see that this contribution is an odd function of the magnetic flux running through the string's core, $\alpha_0$, and it is an even function of the parameter $\tilde{\beta}$. 
\\
On the string's core, $r=0$, $\langle j^\phi\rangle_{\rm{c}}$ is finite and takes the form
\begin{eqnarray}
	\langle j^\phi\rangle_{\rm{c}}|_{r=0}&=&-\frac{16e}{(2\pi)^{5/2}a^5}\sum_{l=1}^{\infty}\cos(2\pi l\tilde{\beta})\mathcal{Z}\big(1+(lL)^2/2w^2\big)\bigg[\sideset{}{'}{\sum}_{k=1}^{[q/2]}(-1)^k\sin(\pi k/q)\sin(2\pi\alpha_0 k)\nonumber \\
	&+&\frac{q}{\pi}\int_{0}^{\infty}dz\frac{\cosh(z)g(q,\alpha_0,2z)}{\cosh(2qz)-\cos(q\pi)}\bigg] \ .
	\label{j-phy-c-req0}
\end{eqnarray}

Combining \eqref{j-phi-cs} and \eqref{j-phy-c}, we may write the total azimuthal current density as
\begin{eqnarray}
\langle j^\phi\rangle&=&-\frac{16e}{(2\pi)^{5/2}a^5}\sideset{}{_*}{\sum}_{l=0}^{\infty}\cos(2\pi l\tilde{\beta})\bigg[\sideset{}{'}{\sum}_{k=1}^{[q/2]}(-1)^k\sin(\pi k/q)\sin(2\pi\alpha_0 k)\mathcal{Z}(u_{lk})\nonumber \\
&+&\frac{q}{\pi}\int_{0}^{\infty}dz\frac{\cosh(z)g(q,\alpha_0,2z)}{\cosh(2qz)-\cos(q\pi)}\mathcal{Z}(u_{lz})\bigg] \ ,
\label{j-phy-c-total}
\end{eqnarray}
where the asterisk sign in the summation over $l$ in \eqref{j-phy-c-total} indicates that the term $l=0$ must be halved.

For a massless charged fermion field we have $\nu_1=\nu_2=1/2$. Thus, taking \eqref{function-z-m0} into \eqref{j-phy-c-total}, the total azimuthal current is simplified and takes the form
\begin{eqnarray}
\langle j^\phi\rangle&=&-\frac{12e}{\pi^2L^5}\bigg(\frac{w}{a}\bigg)^5\bigg[\sideset{}{'}{\sum}_{k=1}^{[q/2]}(-1)^k\sin(\pi k/q)\sin(2\pi\alpha_0 k)V(\tilde{\beta},\rho_k)\nonumber \\
&+&\frac{q}{\pi}\int_{0}^{\infty}dz\frac{\cosh(z)g(q,\alpha_0,2z)}{\cosh(2qz)-\cos(q\pi)}V(\tilde{\beta},\tau(z))\bigg] \ ,
\label{j-phy-massless-case}
\end{eqnarray}
where we have introduced the function
\begin{eqnarray}
V(\tilde{\beta},x)=\sideset{}{_*}{\sum}_{l=0}^{\infty}\frac{\cos(2\pi \tilde{\beta}l)}{(l^2+x^2)^{5/2}} \ ,
\label{func-v}
\end{eqnarray}
in the integrand of \eqref{j-phy-massless-case}, with the corresponding arguments defined as
\begin{equation}
\rho_k=\frac{2r\sin(\pi k/q)}{L} \ \ \ , \ \ \ \tau(z)=\frac{2r\cosh(z)}{L}.
\label{var-2}
\end{equation}
The summation over $l$ can be developed with the help of \cite{Grad}. So, after some intermediate steps, we obtain
\begin{eqnarray}
V(\tilde{\beta},x) =\frac{\pi ^{2}\cosh (2\pi \tilde{\beta}x)}{4x^{2}\sinh
	^{2}(\pi x)}+\pi \frac{\cosh [\pi (1-2\tilde{\beta})x]+2\pi \tilde{\beta}x\sinh [\pi
	(1-2\tilde{\beta})x]}{4x^{3}\sinh (\pi x)} \ ,  \label{Cbet1}
\end{eqnarray}
for $0\leqslant \tilde{\beta}\leqslant 1$.

At large lengths of the compact extra dimension, $L/w\gg1$, and for $ma\ge1$ we have by plugging \eqref{function-z-gg-1} into \eqref{j-phy-c}, the following expression:
\begin{eqnarray}
	\langle j^\phi\rangle_{\rm{c}}&\approx&-\frac{16\nu_2(\nu_2+1)e}{\pi^2a^5}\bigg(\frac{w}{L}\bigg)^{2\nu_2+4}\sum_{l=1}^{\infty}\cos(2\pi l\tilde{\beta})\Bigg\{\sideset{}{'}{\sum}_{k=1}^{[q/2]}\frac{(-1)^k\sin(\pi k/q)\sin(2\pi\alpha_0 k)}{\big(l^2+\rho_k^2\big)^{\nu_2+2}}\nonumber \\&+&
	\frac{q}{\pi}\int_{0}^{\infty}dz\frac{\cosh(z)g(q,\alpha_0,2z)}{[\cosh(2qz)-\cos(q\pi)]\big[l^2+\tau^2(z)\big]^{\nu_2+2}}\Bigg\} \ .
	\label{j-phy-c-asymp-part}
\end{eqnarray}
From the above equation we note that the azimuthal current induced by the compactness starts to fade for large lengths of the compact dimension, and therefore, in this regime, the total azimuthal current is dominated by the string induced contribution.

To conclude this section, we turn our focus to the consideration of the Minkowskian asymptotic limit. In this case, we consider $a\rightarrow\infty$ for fixed value of the coordinate $y$. As a consequence $ma\gg1$ and the coordinate $w$ goes like $w\approx a+y$, such that the function $\mathcal{Z}(u)$ assumes the following form: \cite{Bellucci}
\begin{eqnarray}
\mathcal{Z}(u)=2m^{5/2}a^5\frac{K_{5/2}(mu)}{u^{5/2}} \ ,
\label{z-func-Mink}
\end{eqnarray}
where $K_\nu(x)$ is the Macdonald function. Taking this result into \eqref{j-phy-c-total}, after a few simplifications, we get the following expression:
\begin{eqnarray}
	\langle j^\phi\rangle_{\text{M}}&\approx&-e\bigg(\frac{2m}{\pi L}\bigg)^{5/2}\sideset{}{_*}{\sum}_{l=0}^{\infty}\cos(2\pi l\tilde{\beta})\Bigg\{\sideset{}{'}{\sum}_{k=1}^{[q/2]}(-1)^k\sin(\pi k/q)\sin(2\pi\alpha_0 k)\frac{K_{5/2}(mL\sqrt{l^2+\rho_k^2})}{(l^2+\rho_k^2)^{5/4}}\nonumber \\
	&+&\frac{q}{\pi}\int_{0}^{\infty}dz\frac{\cosh(z)g(q,\alpha_0,2z)}{\cosh(2qz)-\cos(q\pi)}\frac{K_{5/2}(mL\sqrt{l^2+\tau^2(z)})}{[l^2+\tau^2(z)]^{5/4}}\Bigg\} \ .
	\label{j-phy-t-Mink}
\end{eqnarray}
\section{Induced current along the compactified extra dimension}
\label{sec4}
In this section, we want to analyze the VEV of the induced current density along the compactified axis, namely the axial current. As we  are going to see, this VEV goes to zero in the limit $L\rightarrow0$. 

The current density induced in the compactified extra dimension is calculated by 
\begin{equation}
\langle j^z\rangle=\frac{e}{2}\sum_{\sigma}[\bar{\psi}^{(-)}_\sigma\gamma^z\psi^{(-)}_\sigma-\bar{\psi}^{(+)}_\sigma\gamma^z\psi^{(+)}_\sigma] \ .
\end{equation}
Plugging \eqref{positive-energy-wfunc} and \eqref{negative-energy-wfunc} into the above expression, we obtain
\begin{eqnarray}
	\langle j^z\rangle&=&\frac{qew^6}{8\pi La^5}\int_{0}^{\infty}d\lambda\int_{0}^{\infty}dp\sum_{s=\pm1}\sum_{j
	}\sum_{l=-\infty}^{\infty}\frac{\lambda^2p^2}{E\kappa_{s}\sqrt{\tilde{k}_{l}^2+p^2}}s[J^2_{\nu_1}(pw)-\kappa_s^2J^2_{\nu_2}(pw)]\nonumber\\&\times&\big[b_s^{(+)}J^2_{\beta_{j}+\epsilon_{j}}(\lambda r)-b_s^{(-)}J^2_{\beta_{j}}(\lambda r)\big] \ .
\end{eqnarray}
Summing over $s$, we obtain
\begin{eqnarray}
\langle j^z\rangle&=&-\frac{qew^6}{4\pi La^5}\int_{0}^{\infty}d\lambda\int_{0}^{\infty}dp\sum_{j
}\sum_{l=-\infty}^{\infty}\lambda p\bigg\{\frac{\tilde{k}_l}{E}[J^2_{\nu_1}(pw)+J^2_{\nu_2}(pw)]\nonumber\\&\times&\big[J^2_{\beta_{j}}(\lambda r)+J^2_{\beta_{j+\epsilon_{j}}}(\lambda r)\big]+[J^2_{\nu_1}(pw)-J^2_{\nu_2}(pw)]\big[J^2_{\beta_{j}}(\lambda r)-J^2_{\beta_{j+\epsilon_{j}}}(\lambda r)\big]\bigg\} \ .
\end{eqnarray}
 Note that this expression is divergent due to the second term inside the integrand, and therefore some regularization procedure is necessary. We shall assume that a cutoff function is introduced in the aforementioned divergent term without explicitly writing it, since the explicit form of this function is not relevant for this discussion. Moreover, using the Abel-Plana formula given in \eqref{abel-plana}, we can see that this term vanishes. 
 
 For the remaining term, on the other hand, we take $g(u)=2\pi u/L$ and the expression given in \eqref{fu-function} for the function $f(u)$. We can see that the first integral on right-hand side of \eqref{abel-plana} is zero, because $g(u)$ is an odd function. Thus, the only contribution for the axial current comes from the second integral in \eqref{abel-plana}
\begin{eqnarray}
\langle j^z\rangle&=&-\frac{iqew^6}{4\pi^2 a^5} \int_{0}^{\infty}dpp[J^2_{\nu_1}(pw)+J^2_{\nu_2}(pw)]\sum_{j
}\int_{0}^{\infty}d\lambda\lambda\big[J^2_{\beta_{j}}(\lambda r)+J^2_{\beta_{j+\epsilon_{j}}}(\lambda r)\big]\nonumber\\&\times&\int_{\sqrt{\lambda^2+p^2}}^{\infty}\frac{dxx}{\sqrt{x^2-\lambda^2-p^2}}\bigg(\frac{1}{e^{Lx+2\pi i\tilde{\beta}}-1}-\frac{1}{e^{Lx-2\pi i\tilde{\beta}}-1}\bigg) \ ,
\end{eqnarray}
where we have introduced the variable $x=2\pi u/L$. As before, the next step is to use the expansion $(e^u-1)^{-1}=\sum_{l=0}^{\infty}e^{-lu}$, in the above expression, and with the help of \cite{Grad} the integral over $x$ can be performed, with the result given in terms of the Macdonald function, $K_1(z)$. 
\begin{eqnarray}
\langle j^z\rangle&=&-\frac{qew^6}{2\pi^2 a^5}\sum_{l=1}^{\infty}\sin(2\pi l\tilde{\beta}) \int_{0}^{\infty}dpp[J^2_{\nu_1}(pw)+J^2_{\nu_2}(pw)]\nonumber\\&\times&\sum_{j
}\int_{0}^{\infty}d\lambda\lambda\big[J^2_{\beta_{j}}(\lambda r)+J^2_{\beta_{j+\epsilon_{j}}}(\lambda r)\big]\sqrt{\lambda^2+p^2}K_1(lL\sqrt{\lambda^2+p^2}) \ .
\end{eqnarray}
Using the integral representation for the Macdonald function already given in \eqref{Macdonald-int-represent} and the fact that $K_\nu(z)=K_{-\nu}(z)$, we can write
\begin{eqnarray}
\langle j^z\rangle&=&-\frac{qew^6}{2\pi^2 La^5}\sum_{l=1}^{\infty}\frac{\sin(2\pi l\tilde{\beta})}{l} \int_{0}^{\infty}dse^{-s}\int_{0}^{\infty}dppe^{-(lL)^2p^2/4s}[J^2_{\nu_1}(pw)+J^2_{\nu_2}(pw)]\nonumber\\&\times&\sum_{j
}\int_{0}^{\infty}d\lambda\lambda e^{-(lL)^2\lambda^2/4s}\big[J^2_{\beta_{j}}(\lambda r)+J^2_{\beta_{j+\epsilon_{j}}}(\lambda r)\big] \ .
\end{eqnarray}
Performing the integrals over $\lambda$ and $p$ with the help of \cite{Grad}, we obtain
\begin{eqnarray}
\langle j^z\rangle&=&-\frac{qew^6L}{4\pi^2 a^5r^6}\sum_{l=1}^{\infty}l\sin(2\pi l\tilde{\beta}) \int_{0}^{\infty}dyy^2e^{-[1+(lL)^2/2r^2+w^2/r^2)]y}\nonumber\\&\times&\bigg[I_{\nu_1}\bigg(\frac{w^2}{r^2}y\bigg)+I_{\nu_2}\bigg(\frac{w^2}{r^2}y\bigg)\bigg]\big[\mathcal{I}(q,\alpha_0,y)+\mathcal{I}(q,-\alpha_0,y)\big] \ ,
\end{eqnarray}
where we have introduced the new
 variable $y=2sr^2/(lL)^2$.

Using again \eqref{j-sum} for the function $\mathcal{I}(q,\alpha_0,y)$, we can evaluate the integral over $y$, obtaining
 \begin{eqnarray}
 \langle j^z\rangle&=&-\frac{8eL}{(2\pi)^{5/2} a^5}\sum_{l=1}^{\infty}l\sin(2\pi l\tilde{\beta})\bigg[\sideset{}{'}{\sum}_{k=0}^{[q/2]}(-1)^k\cos(\pi k/q)\cos(2\pi\alpha_0 k)\mathcal{Z}(u_{lk})\nonumber\\&+&\frac{q}{\pi}\int_{0}^{\infty}dz\frac{\sinh(z)h(q,\alpha_0,2z)}{\cosh(2qz)-\cos(q\pi)}\mathcal{Z}(u_{lz})\bigg] \ ,
 \label{axial-curr-full}
 \end{eqnarray}
 where we have introduced the function
 \begin{eqnarray}
	h(q,\alpha_0,z)&=&\cos[q\pi(1/2+\alpha_0)]\sinh[q(1/2-\alpha_0)z]\nonumber\\&+&\cos[q\pi(1/2-\alpha_0)]\sinh[q(1/2+\alpha_0)z] \ .
 \end{eqnarray}
\begin{figure}[!htb]
	\begin{center}
		\includegraphics[scale=0.5]{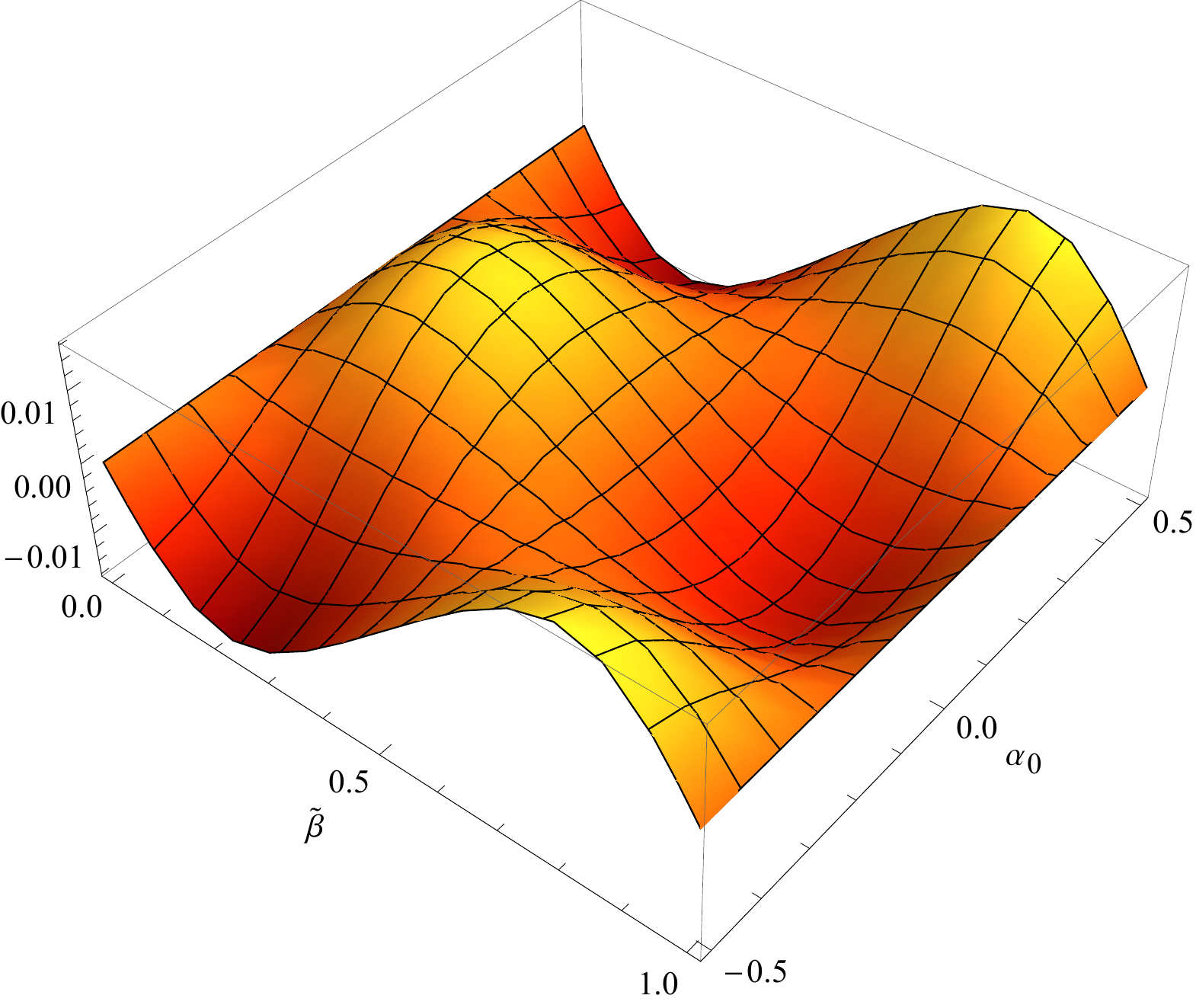}
		\caption{The VEV of the axial current induced by the compactification in Eq.\eqref{axial-curr-kdiff0} is plotted, in units of $a^{5}e^{-1}$,  for $q=2.5$, $r/w=1$ and $ma=5$.}
		\label{fig4}
	\end{center}
\end{figure}\\
Note that the prime on the sign of the sum means that the term with $k=0$ should be halved. The VEV in \eqref{axial-curr-full} may be also represented as
\begin{equation}
	\langle j^z\rangle=\langle j^z\rangle^{(0)}+\langle j^z\rangle^{(q,\alpha_0)} \ ,
	\label{axial-decomp}
\end{equation}
The first term, $\langle j^z\rangle^{(0)}$,  does not depend on $\alpha_0$ and $q$, and it corresponds to the current density induced only by compactification in AdS spacetime. For this contribution one has
\begin{equation}
	\langle j^z\rangle^{(0)}=-\frac{4qeL}{(2\pi)^{5/2}a^5}\sum_{l=1}^{\infty}l\sin(2\pi l\tilde{\beta})\mathcal{Z}(u_{l0}) \ ,
	\label{axial-curr-k0}
\end{equation}
where $u_{l0}$ is given in the first relation of \eqref{var} for $k=0$. 
The second term in \eqref{axial-decomp} is given by
\begin{eqnarray}
	\langle j^z\rangle^{(q,\alpha_0)}&=&-\frac{8eL}{(2\pi)^{5/2} a^5}\sum_{l=1}^{\infty}l\sin(2\pi l\tilde{\beta})\bigg[\sideset{}{'}{\sum}_{k=1}^{[q/2]}(-1)^k\cos(\pi k/q)\cos(2\pi\alpha_0 k)\mathcal{Z}(u_{lk})\nonumber\\&+&\frac{q}{\pi}\int_{0}^{\infty}dz\frac{\sinh(z)h(q,\alpha_0,2z)}{\cosh(2qz)-\cos(q\pi)}\mathcal{Z}(u_{lz})\bigg] \ .
	\label{axial-curr-kdiff0}
\end{eqnarray}
From \eqref{axial-curr-kdiff0} one can see that this contribution for the axial current density vanishes, for the special cases of untwisted and twisted field configuration, in the absence of magnetic flux enclosed by the compactified extra dimension, and more in general for half-integer values of $\tilde{\beta}$. Another feature of this component is the periodicity regarding to the magnetic fluxes, being it an even function of the $\alpha_0$ and an odd function of the magnetic flux enclosed by compactified axis for the untwisted field case. In Fig.\ref{fig4} is presented a plot of \eqref{axial-curr-kdiff0} in terms of the $\alpha_0$ and $\tilde{\beta}$, for values of $r/w=1$, $ma=5$ and $q=2.5$. In figure \ref{fig5and6} we plot \eqref{axial-curr-kdiff0} for $\alpha_0=0$ (left panel) and $\alpha_0=0.35$ (right panel) as a function of $\tilde{\beta}$ for different values of $q$. As we can observe, the magnetic flux along the cosmic string can invert the direction of the current density along the axial axis.

For a massless spinor field, we have $\nu_{1}=\nu_{2}=1/2$. Thus, by plugging \eqref{function-z-m0} into \eqref{axial-curr-full}, we find
\begin{eqnarray}
	\langle j^z\rangle&=&-\frac{6e}{\pi^2 L^4}\bigg(\frac{w}{a}\bigg)^5\bigg[\sideset{}{'}{\sum}_{k=0}^{[q/2]}(-1)^k\cos(\pi k/q)\cos(2\pi\alpha_0 k)G(\tilde{\beta},\rho_k)\nonumber\\&+&\frac{q}{\pi}\int_{0}^{\infty}dz\frac{\sinh(z)h(q,\alpha_0,2z)}{\cosh(2qz)-\cos(q\pi)}G(\tilde{\beta},\tau(z))\bigg] \ ,
	\label{axial-curr-massless-case}
\end{eqnarray}
where we have defined a new function
\begin{equation}
	G(\tilde{\beta},x)=\sum_{l=1}^{\infty}\frac{l\sin(2\pi l\tilde{\beta})}{(l^2+x^2)^{5/2}} \ .
\end{equation}
The summation on the quantum number $l$ can be developed with the help of \cite{Grad}. The result is
	\begin{eqnarray}
	G({\tilde{\beta}},x)=-\frac{\pi^2}{4x}\frac{[\sinh(2\pi{\tilde{\beta}}x)-
		2{\tilde\beta}\sinh(\pi x)\cosh[\pi x(1-2{\tilde\beta})]]}{\sinh^2(\pi x)} \ .
	\label{sum-sin1}
	\end{eqnarray} 
	Clearly we can see that $G({\tilde{\beta}},x)$ vanishes for ${\tilde{\beta}}=0, \ 1/2 , \ 1$.

Now let us turn our attention to some special asymptotic limits of \eqref{axial-curr-full}. Taking $w$ fixed and large values of lengths for the compact dimension, $L/w\gg1$, the total axial current is expressed as
\begin{eqnarray}
	\langle j^z\rangle&\approx&-\frac{8\nu_2(\nu_2+1)eL}{\pi^2 a^5}\bigg(\frac{w}{L}\bigg)^{2\nu_2+4}\sum_{l=1}^{\infty}l\sin(2\pi l\tilde{\beta})\Bigg\{\sideset{}{'}{\sum}_{k=0}^{[q/2]}(-1)^k\frac{\cos(\pi k/q)\cos(2\pi\alpha_0 k)}{\big(l^2+\rho_k^2\big)^{\nu_2+2}}\nonumber\\&+&\frac{q}{\pi}\int_{0}^{\infty}dz\frac{\sinh(z)h(q,\alpha_0,2z)}{[\cosh(2qz)-\cos(q\pi)]\big[l^2+\tau^2(z)\big]^{\nu_2+2}}\Bigg\} \ ,
	\label{axial-curr-full-asymp}	
\end{eqnarray}
which is valid only for values of $ma\ge1/2$.
\begin{figure}[!htb]
	\begin{center}
		\includegraphics[scale=0.45]{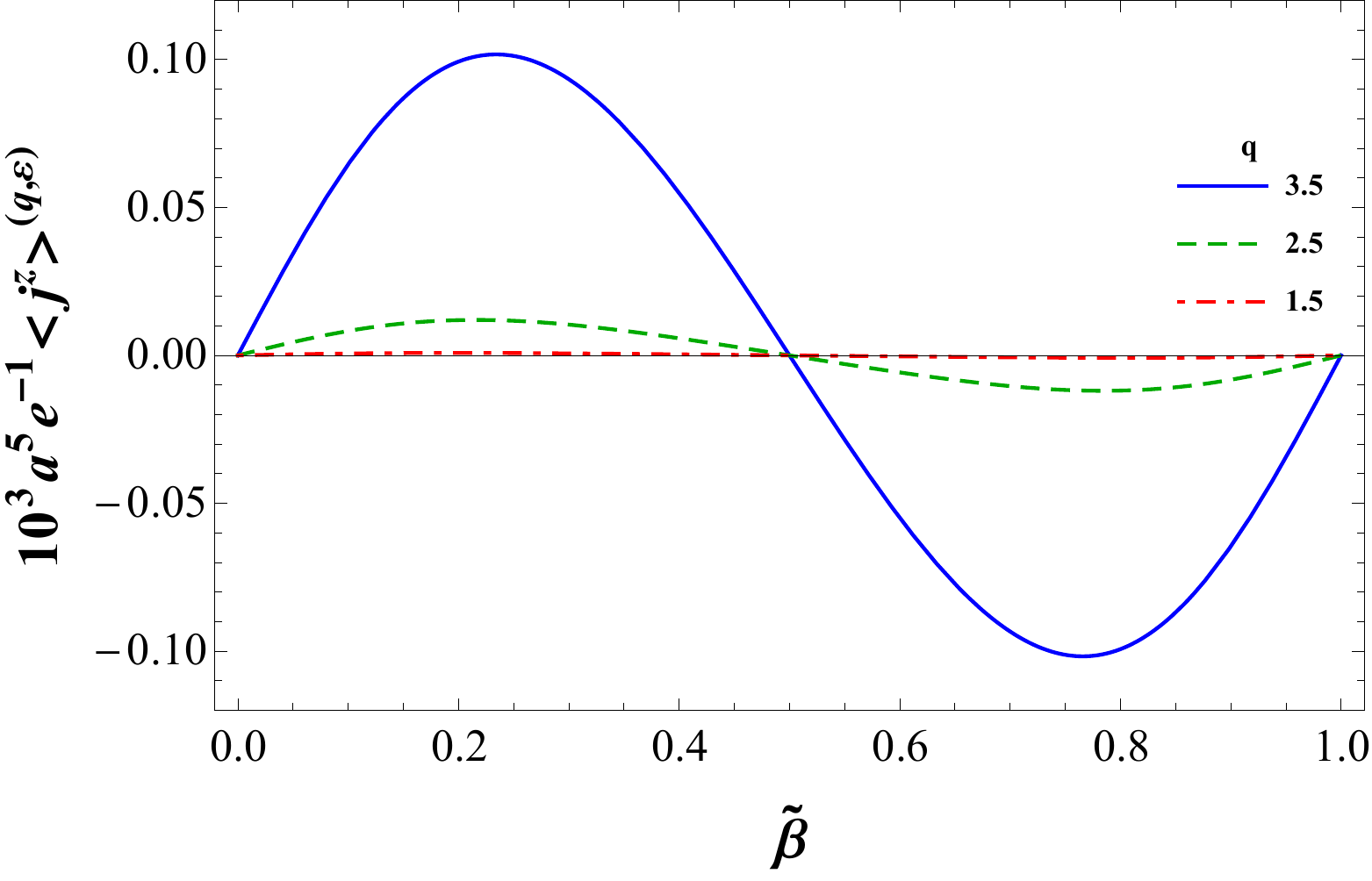}
		\quad
		\includegraphics[scale=0.45]{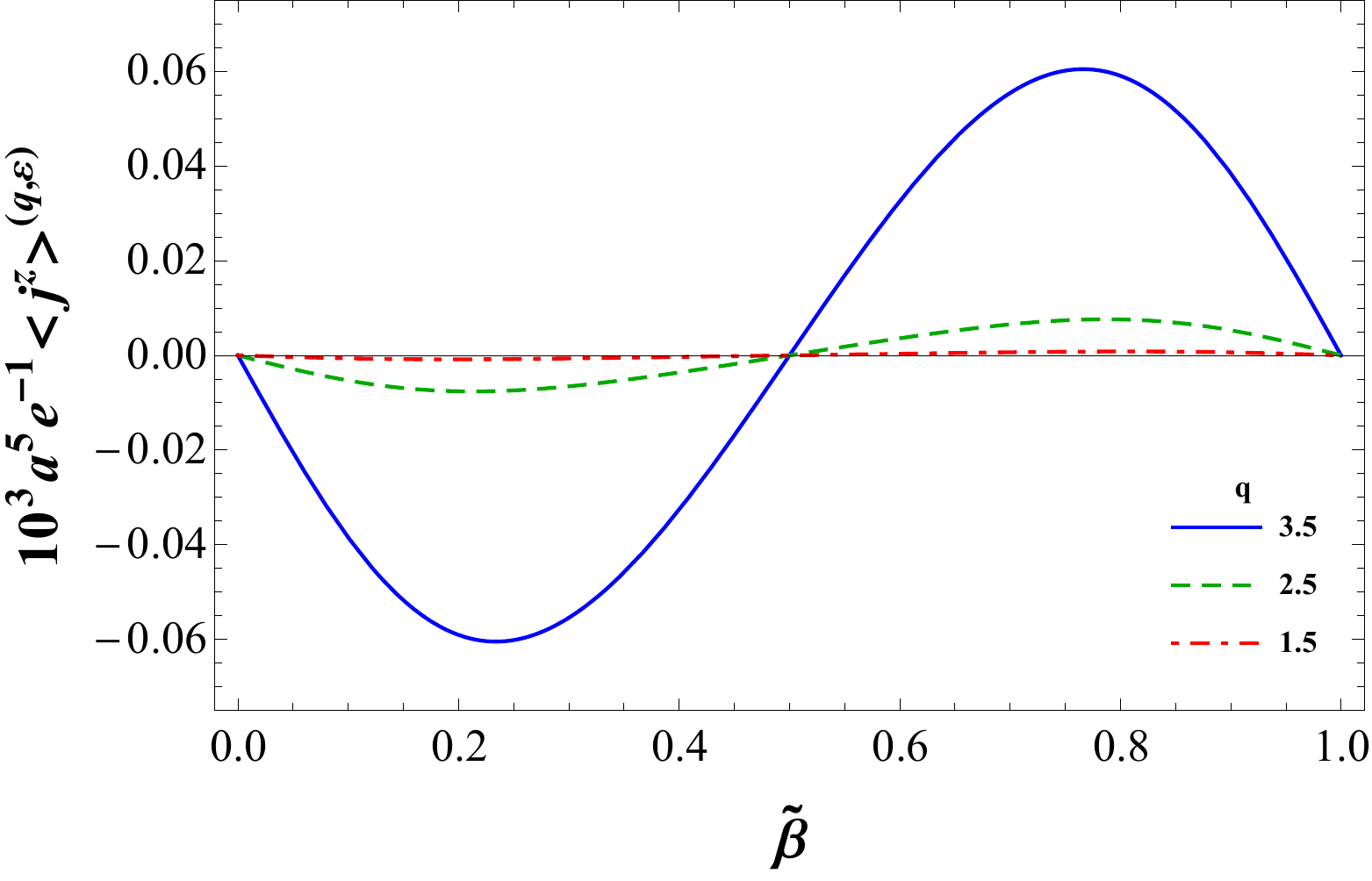}
		\caption{The VEV of the axial current induced by the compactification in Eq.\eqref{axial-curr-kdiff0} is plotted, in units of $a^{5}e^{-1}$, as a function of $\tilde{\beta}$ for $q=2.5$, $r/w=1$ and $ma=5$. In the left plot, we consider $\alpha_0=0$, while in the right plot, we take $\alpha_0=0.35$.}
		\label{fig5and6}
	\end{center}
\end{figure}\\ 

Another very interesting asymptotic behavior comes from the Minkowskian limit. Making use again of the corresponding $\mathcal{Z}(u)$ function in this limit given by \eqref{z-func-Mink}, and taking it into \eqref{axial-curr-full}, the total axial current density is expressed as
\begin{eqnarray}
	\langle j^z\rangle_\text{M}&\approx&-\frac{2^{3/2}m^{5/2}e}{\pi^{5/2} L^{3/2}}\sum_{l=1}^{\infty}l\sin(2\pi l\tilde{\beta})\Bigg\{\sideset{}{'}{\sum}_{k=0}^{[q/2]}(-1)^k\cos(\pi k/q)\cos(2\pi\alpha_0 k)\frac{K_{5/2}\big(mL\sqrt{l^2+\rho_k^2}\big)}{(l^2+\rho_k^2)^{5/4}}\nonumber\\&+&\frac{q}{\pi}\int_{0}^{\infty}dz\frac{\sinh(z)h(q,\alpha_0,2z)}{\cosh(2qz)-\cos(q\pi)}\frac{K_{5/2}\big(mL\sqrt{l^2+\tau^2(z)}\big)}{[l^2+\tau^2(z)]^{5/4}}\Bigg\} \ .
	\label{axial-curr-Mink-limit}
\end{eqnarray}
\section{Conclusion}\label{sec5}
In this paper, we have considered a massive charged  fermionic field and investigated the induced current densities in a 5-dimensional AdS space-time, assuming the presence of a cosmic string having a magnetic flux running along its axis. Additionally, we admit that an extra dimension is compactified to a circle of perimeter $L$ and that, in addition, a magnetic flux enclosed by the compactified axis exists. In order to develop this analysis, we calculated the complete set of fermionic wave functions of positive and negative frequency obeying quasi-periodicity condition, with an arbitrary phase, $\beta$, along the compact dimension. To calculate the VEV of the fermionic induced current, we have made use of the sum mode formula given in \eqref{curr-dens-formula} and the positive and negative fermionic modes given in \eqref{positive-energy-wfunc} and \eqref{negative-energy-wfunc}, respectively. In our analysis we proved that only azimuthal and axial current densities take place. These VEVs are periodic functions of the magnetic flux running along the string, $\alpha_0$, and the parameter $\tilde{\beta}$. This is an Aharonov-Bohm-like effect.

By using the Abel-Plana summation formula given in \eqref{abel-plana}, we were able to decompose the VEV of the total azimuthal current density given in \eqref{j-phy-c-total} into two parts, one purely associated to the cosmic string (corresponding to $l=0$) and another related to the compactification ($l\neq0$). The string contribution is given in \eqref{j-phi-cs} and it is an odd function of the magnetic flux, $\alpha_0$, with its periodic behavior displayed in Fig.\ref{fig1}. The analysis of the asymptotic behaviors was made for points close and far away from the string, and it is presented in \eqref{asymp-close-cs} and \eqref{asymp-far-away-cs}, respectively. In the former, $r/w\ll1$, $\langle j^\phi\rangle_{\rm{cs}}$ diverges on the string as $(w/r)^5$; in the latter, $r/w\gg1$, the current density starts to fade away with its decay strength depending on the mass of the fermion. These behaviors are graphically exhibited in figure \ref{fig2}.

As to the azimuthal current induced by compactification given in Eq.\eqref{j-phy-c}, we note that it is an odd function of the magnetic flux along the string's core, $\alpha_0$, and it is an even function of the parameter, $\tilde{\beta}$, with a period equal to the quantum flux $\Phi_{0}$. This periodic behavior can be observed in Fig.\ref{fig3}. In particular, for $\beta=0$, $\langle j^\phi\rangle_{\rm{c}}$ becomes an even function of the magnetic flux enclosed by the compact dimension.  Besides, it is finite on the string's core, and its asymptotic expression for large lengths of the compactified dimension is presented in \eqref{j-phy-c-asymp-part}. The total azimuthal current density is calculated for a massless spinor field, and it is given in \eqref{j-phy-massless-case}. In \eqref{j-phy-t-Mink} we present the Minkowskian asymptotic limit for the total induced azimuthal current density.

The presence of compactification also gives rise to an induced axial current density, and it is presented in closed form in \eqref{axial-curr-full}. This contribution can be separated into two components. The first one, given in \eqref{axial-curr-k0}, comes from the term $k=0$ and does not depend on the radial distance $r$, neither in $q$ or $\alpha_0$. The second one, given by \eqref{axial-curr-kdiff0}, however, depends on the aforementioned parameters, and it is an even function of $\alpha_0$ and an odd function of the parameter $\tilde{\beta}$, with this periodic behavior displayed in figure \ref{fig4}. For $\alpha_0=0$ and $\alpha_0=0.35$ we have plotted Eq.\eqref{axial-curr-kdiff0} in Fig.\ref{fig5and6}. The plots show that the magnetic flux along the string's core has the effect of inverting the direction of the current. The expression for a massless field is presented in \eqref{axial-curr-massless-case}. For large lengths of the compact dimension, an asymptotic expression is given in \eqref{axial-curr-full-asymp} and shows that the induced axial current vanishes when $L\rightarrow\infty$. Finally, the Minkowskian asymptotic limit is calculated for the total axial current density, and it is given in \eqref{axial-curr-Mink-limit}.

In conclusion, we have showed in this paper that the fermionic current densities induced in the vacuum are not only generated by the magnetic fluxes, but also from the nontrivial topology  of the compactified extra dimension introduced by the arbitrary phase in the quasi-periodic condition. As it was showed graphically, the angular deficit produced on the background geometry by the cosmic string plays an important role on the intensity and behavior of the VEVs analysed.
\section*{Acknowledgments}
Discussions and critical reading of the manuscript by Aram Saharian are gratefully acknowledeged. This study was financed in part by the Coordena\c c\~ao de Aperfei\c coamento de Pessoal de N\'ivell Superior - Brasil (CAPES) - Finance Code 001.

\end{document}